\documentstyle [aps,pre,epsf,preprint] {revtex}

\tightenlines
\begin{document}
%\draft

%\setlength{\baselineskip}{18pt}
%-----------------------------------------------------------
% 1st page
%-----------------------------------------------------------
%%%%%%%%%%%%%%%%%%%%%%%%%%%%%%%%%%%%%%%%%%%%%%%%%%%%%%%%%%%%%%%
\title {{\bf The Lamellar-Disorder Interface:}\\
       {\bf One-Dimensional Modulated Profiles}}
%%%%%%%%%%%%%%%%%%%%%%%%%%%%%%%%%%%%%%%%%%%%%%%%%%%%%%%%%%%%%%
%{\centerline {{\bf {Interface in Modulated Phase Systems}}}}
%\vskip 2cm
\author{ {Simon Villain-Guillot
{\footnote{Permanent address: \\
    Groupe de Physique Statistique   \\
    Universit\'e de Cergy-Pontoise  \\
    BP 8428, 95806 Cergy-Pontoise Cedex, France}}
and David Andelman}  \\
    {\em School of Physics and Astronomy }  \\
    {\em Raymond and Beverly Sackler Faculty of Exact Sciences}\\
    {\em Tel-Aviv University} \\
    {\em Ramat-Aviv 69978, Tel-Aviv, Israel}}
%\date{\today}
\maketitle
\begin{abstract}
We study interfacial behavior of a lamellar (stripe) phase
coexisting with a disordered phase.
Systematic analytical expansions are obtained for the interfacial
profile in the vicinity of a tricritical point. They are characterized
by a wide interfacial region involving a large number of lamellae.
Our analytical results apply to systems with one dimensional symmetry
in true thermodynamical equilibrium and are of relevance
to metastable interfaces between lamellar and disordered phases
in two and three dimensions. In addition, 
good agreement is found with 
numerical minimization schemes of the full
free energy functional having the same one dimensional symmetry.
The interfacial energy for the lamellar to disordered
transition is obtained in accord with mean field scaling laws of
tricritical points.
\end{abstract}

%------------------------------------------------------
\newpage
%%%%%%%%%%%%%%%%%%%%%%%%%%%%%%%%%%%%%%%%%%%%%%%%%%%%%%%%%%%%%%%
\section{Introduction}
%%%%%%%%%%%%%%%%%%%%%%%%%%%%%%%%%%%%%%%%%%%%%%%%%%%%%%%%%%%%%%%

\noindent

Some equilibrium phases of matter exhibit spatial modulations
in one or more of their local properties (local order parameter)
and hence are called {\em modulated phases} \cite{SC}. 
Modulated phases can be found in a variety of 
physical and chemical systems such as
diblock copolymers, magnetic garnet films,
ferrofluids, adsorbates on surfaces of metals, superconductors 
(Type I) and semiconductors.
Although these  systems all have  
spatial modulations in their order parameter, 
the microscopic origin
of these modulations is very different.

Usually modulated phases manifest either a one dimensional packing
of sheets (the so-called lamellar or stripe phase), 
two-dimensional packing of cylinders with an hexagonal symmetry
or a cubic packing of spheres or other, more complex,
bi-continuous super-structures with cubic symmetry.
The origin 
of the preferred 
length scale of the modulations can be 
explained in many cases by a competition between two
type of interactions: 
a short-range interaction which tends to make the
system more homogeneous 
together with a long-range one, or a non-local one, which 
prefers proliferation of domain walls. For example, in the magnetic
films, the dipole-dipole interaction has the latter effect.

These two opposing interactions are responsible in the above systems for
the appearance of spatially modulated phases having an optimal non-zero
wave-vector. Depending on the range of the thermodynamical variables,
instead of having the usual segregation of the system into two homogeneous
phases (such as coexisting liquid and gas phases, or two coexisting
polymer phases), the system self-organizes into alternating domains. 
Those domains are associated with spatially modulated order parameter
(polarization, magnetization, relative concentration), and can be regarded
as unit cells of super-structures of lamellar, hexagonal or cubic
symmetry. Depending on the system in mind, the domain length scale can be
as small as one hundred angstroms or as large as centimeters \cite{SC}. In
contrast to periodic crystalline structures, the periodicity in modulated
phases can be tuned by varying external fields such as pressure,
temperature and electric or magnetic fields. 

In this paper we would like to present analytical results related to
interfaces in modulated phases. When the modulations are relatively weak
an analytical expansion of the interfacial profile can be obtained
systematically. In particular, we calculate the profile of the
lamellar-to-disordered interface for a model free energy
where spatial modulations are restricted along {\it one dimension}. 
Physical systems can create
more complicated modulations in two- and three dimensions. In many
cases, the lamellar-to-disordered interface addressed here
can be shown to be only {\it metastable} \cite{netz,FB}
and not in true thermodynamical equilibrium. 
Nevertheless, their systematic study can be useful to understand
nucleation and growth in e.g., polymeric systems with slow dynamics.
In addition, the rather simplified lamellar-disordered interface
can be generalized to other situations.

Our analytical results are valid in the vicinity of a tricritical
point which occurs for asymmetric profiles in our strictly
one-dimensional model. Close to the tricritical point 
it is  possible to decouple the two characteristic length scales
which appears in the model~: the period associated with the modulated
phase and the width of the interface. Thus, it is possible to compute
analytically the leading terms of the interfacial energies. 

The obtained results presented here are in good agreement with recent work
by Netz et al \cite{netz} where numerical minimization of a similar free
energy functional yielded detailed structure of interfaces between various
{\it two-dimensional} modulated phases. 

%\newpage

%%%%%%%%%%%%%%%%%%%%%%%%%%%%%%%%%%%%%%%%%%%%%%%%%%%%%%%%%%%%
\section{The model and its phase diagram}
%%%%%%%%%%%%%%%%%%%%%%%%%%%%%%%%%%%%%%%%%%%%%%%%%%%%%%%%%%%%

\noindent

A phenomenological free energy
$F$ similar to the one introduced in  Ref. \cite{netz}
is used hereafter in order to allow us
to study one-dimensional profiles between
lamellar and disordered phases.
\begin{equation}
 F =2 (\nabla ^2 \phi ) ^ 2 - 2 (\nabla \phi)^2 +
\frac {t}{2} \phi^2 + \frac {1}{4} \phi^4 \label{free}
 \end{equation}

\noindent
This mean-field free energy is a modified Ginzburg-Landau (GL) expansion 
in the order parameter $\phi(x)$, which varies
between $\phi=1$ for A-rich domains to 
$\phi=0$ for B-rich domains. 
The last two terms in $F$
are the usual terms appearing in the Landau expansion
with $t=(T-T_c)/T_c$ being the reduced temperature and the
coefficient of the fourth order term is chosen to be $1/4$
without loss of generality. 
For an order parameter without spatial modulations they define our
disorder free energy, $F_{_D}$ (within the Landau expansion):
\begin{equation}
F_{_D}=\frac {t}{2} \phi^2 + \frac {1}{4} \phi^4
\label{free_D}
\end{equation}
\noindent
The main deviation from the 
regular GL expansion can be seen in the first two
terms of (\ref{free}). 
The coefficient of the gradient square term is negative 
(favoring modulations), whereas the coefficient
of the Laplacian square term is positive
(preventing strong modulations) and playing
the role of a homogenizing interaction.
An illustrative example can be
a diblock copolymer melt where each polymer
is composed of two blocks,
A and B, which would like to phase separate. However, since the A and
B chains are chemically linked forming one polymer chain, 
the system can only 
undergo a micro-phase separation resulting in a spatial modulation
in the local A/B relative concentration.

In order to construct the phase diagram, 
we compare the mean-field free energy 
$F_{_D}$ associated with the disordered
phase ($\phi=const.$)  with the free
energy $F_{_L}$ of a phase described by a modulating  
$\phi(\vec{r})$.  Our approach involves several simplifying assumptions:
In the vicinity of a critical point
(weak segregation limit), an analytical
expansion of the free energy in powers
of the small order parameter (Landau expansion)
can  be justified. Since 
 a mean-field approximation is employed,
 critical fluctuations corrections  are omitted \cite{BR,GA}. Furthermore,
only the most dominant
modulation mode (the so-called single-mode approximation)
is considered because the amplitude of the fluctuations of other
$q$-modes grows more slowly in this region as was shown by Brazovskii
\cite{BR,GA,GP}. 
Finally, 
only one-dimensional modulations of $\phi(\vec{r})=\phi(x)$ 
along an arbitrary chosen $x$ direction will be explored.
This reduces the problem into an effective one-dimensional problem.
It applies to interfaces where the lamellae are themselves
parallel to the lamellar - disorder interface. We further discuss 
the validity and limitations of our assumptions in the last section.
 
\begin{equation} 
\phi = \phi_o + 2 \phi_q \cos(qx) \label{brazo}
\end{equation} 

\noindent
where $\phi_o$ is the mean spatial order parameter (the average A/B 
relative concentration) and $\phi_q$ is the amplitude of the $q$-mode
modulation.

The most dominant mode is  
obtained by minimizing the free energy with respect to the 
$q$-vector. Within the single mode approximation this leads
to
{$q^*= 1/ \sqrt{2}$} with our choice of simple coefficients
in (\ref{free}). This is the wave-vector characterizing
the periodicity of the
super-structure, $2\pi/q^*$. 
Inserting $\phi(x)$ of the dominant mode from (\ref{brazo})  
into (\ref{free}) we obtain the lamellar free energy

\begin{equation} 
F_{_L} =\frac {t}{2} \phi_o^2 + \frac 14 \phi_o^4 +  
(t-1+3\phi_o^2)\phi_q^2
+ \frac 32 \phi_q^4\label{fl}  
\end{equation} 

\noindent
The disorder free energy $F_{_D}$ (\ref{free_D})
is simply given by setting $\phi_q^2=0$ in
the above expression for $F_{_L}$.

Since this expression 
is symmetric with respect to $\phi _o \to - \phi _o$
only the $\phi_o \ge 0$ region can be
considered without loss of generality. 
We then apply the
variational principle with respect to $\phi_q$. This enables us to
integrate over $\phi_q$ --- one of the two order parameters. 
Since the amplitude of the modulation
minimizing $F_{_L} $ is~:  

\begin{equation} 
\phi_q ^2 = \frac {1-t}{3} - \phi_o^2
\label{second} 
\end{equation} 

\noindent
the necessary condition for the existence of this modulated phase is
$\phi_q^2\ge 0$, hence

 \begin{equation} 
\frac {1-t}{3} - \phi_o^2 \geq 0 \label{conv} 
\end{equation} 

\noindent
Substituting $\phi_q^2$ from (\ref{second}) into (\ref{fl}) 
gives the free energy
for the lamellar phase 

\begin{equation} 
F_{_L} =- \frac {(t-1)^2}{6}+(1-\frac t2) \phi_o^2 -\frac 54 \phi_o^4
\end{equation} 

In the region bounded by inequality (\ref{conv}) and close
to the critical point, the free energy is
a convex function of $\phi_o$.  Therefore, the 
second-order phase transition line separating the order and the 
disordered 
phases is given by:  

\begin{equation} 
\phi_o^2 = \frac {1-t}{3} \label{slope} 
\end{equation} 

\noindent
showing an upward shift of the critical temperature $t_c =1$ from
its usual value for the GL model, $t_c =0$.  A tricritical
point appears at a temperature $t_{tcp} = \frac {3}{4}$, for which 
$ {\delta^2 F_{_L}} / {\delta \phi_o^2}=0$.  
Below this temperature, the free energy is concave; the system 
therefore prefers to phase separate.  Equation (\ref{second}) 
gives the value 
of the order parameter (concentration)  at the tricritical point, 
$ \phi_{_T}^2 = \phi^2 (t_{tcp}) = \frac {1}{12}$.

 In Figs. 1 and 2 the phase diagram is plotted as function of the
reduced temperature -- average concentration ($t$, $\phi_o$), and as
function of the chemical potential -- reduced temperature ($\mu$, $t$),
respectively.  The phase diagram is calculated numerically within the
single mode approximation by comparing the lamellar free energy $F_{_L}$
with the disorder one, $F_{_D}$.  We note that the phase diagram in the
vicinity of the tricritical point resembles that of the BEG
(Blume-Emery-Griffiths) spin one model \cite{BEG}. The BEG model also has
two order parameters and exhibits a similar behavior: a first order phase
transition line meets a second-order one at a tricritical point. 

In the vicinity of the tricritical point (but inside the
two phase region) we are still in the weak segregation
limit. Hence,  a systematic expansion can be 
performed as function of
a new reduced temperature $\varepsilon$, measuring
the distance from the tricritical point:
$\varepsilon = t_{tcp} - t>0$. We then get~:  

\begin{equation} 
F_{_L} = \frac {t}{2} \phi_0^2 +\frac 14 \phi_0 ^4 -
\frac 32 ( \phi_{_T}^2+ \frac { \varepsilon }{3} - \phi_0^2 )^2
\end{equation}
This equation can be expanded in terms of
$\eta_{_D}$ and $\eta_{_L}$, such that 
$ \phi (\varepsilon)  = \phi_{_T} +\eta_{_D} (\varepsilon)$ for the
disordered (isotropic) 
phase and $ \phi (\varepsilon)  = \phi_{_T} +\eta_{_L}(\varepsilon)$ 
for the lamellar
phase. In the two-phase coexistence region, the boundaries are given by
the common tangent method~:  

\begin{equation} 
\mu_{_{D}} =  \mu_{_{L}}=\mu \label{a}  
\end{equation}

\noindent
together with  
\begin{equation} 
G_{_L} \equiv F_{_L} - \mu \phi_{_L} = G_{_D} \equiv
F_{_D} - \mu \phi_{_D} \label{b}
\end{equation} 

\noindent
 where $F_{_L}$ and $ F_{_D}$ are expressed in term of $\eta_{_D}$ and
$\eta_{_L}$, respectively.

\begin{eqnarray}
 F_{_L} &= &\frac {t}{2} \phi_{_T}^2 +\frac 14 \phi_{_T}^4 - {\varepsilon^2
\over 6} + ({ 5 \over 6} + \varepsilon ) \phi_{_T} \eta_{_L} + {\varepsilon\over
2} {\eta_{_L}^2} - 5 \phi_{_T}\eta_{_L}^3 - {5 \over 4 } \eta_{_L}^4
 \\
F_{_D} &= &\frac {t}{2}
\phi_{_T}^2 +\frac 14 \phi_{_T} ^4 +({5 \over 6} - 
\varepsilon ) \phi_{_T} \eta_{_D} + ( 1-\varepsilon)
{\eta_{_D} ^ 2 \over 2} + \phi_{_T} \eta_{_D} ^ 3 + { 1 \over 4 } \eta_{_D} ^ 4 
\end{eqnarray} 

%\noindent
Relation (\ref{a}) enables us to write $\eta_{_D}$ as an 
expansion in $\eta_{_L}$ up to third order in $\varepsilon$~:  

\begin{equation} 
\eta_{_D}=2 \phi_{_T} \varepsilon + \phi_{_T} \varepsilon^2 + 
\varepsilon \eta_{_L} - 15 \phi_{_T} \eta_{_L} ^ 2 
- { 2 \over 3 } \phi_{_T} \varepsilon^3 - 5 \eta_{_L}^3
\end{equation} 

\noindent
and relation (\ref{b}) leads to $\eta_{_L}$ up to third order in $\varepsilon$~:  

\begin{equation} 
 \eta_{_L}=\phi_{_T}\Bigl(-\frac 25 \varepsilon +\frac 1{25} \varepsilon ^2
+\frac{13}{250}\varepsilon^3\Bigr) 
 \end{equation} 

\noindent
Therefore~: 

\begin{equation} 
 \eta_{_D}=\phi_{_T} \Bigl(2\varepsilon +\frac 25 \varepsilon ^2 -\frac {14}{25}
\varepsilon ^3 \Bigr)  \label{pente}
 \end{equation}

In terms of newly defined order parameters: $\phi_\pm
\equiv (\phi_{_D} \pm \phi_{_L})/2$, we get

\begin{eqnarray} 
{\phi_{+}}=\frac 12 ( \phi_{_D} + 
\phi _{_L}) & = & \phi_{_T}+ \frac 12 ( \eta_{_D}+\eta_{_L}) 
 =   \phi_{_T} \Bigl( 1+{ 4 \over 5}\varepsilon +{{11} \over {50}} \varepsilon^2 
-\frac{127}{250} \varepsilon ^3 \Bigr) \nonumber
\\
 {\phi_{-} }=\frac 12 (\phi_{_D} -\phi _{_L}) &=& \frac 12 ( \eta_{_D} -
\eta_{_L})
 = \phi_{_T}\Bigl( {6 \over 5} \varepsilon + {9 \over {50}} \varepsilon^2
-\frac{153}{250} \varepsilon ^3 \Bigr) \label{expansion} \\ \phi _q ^ 2 =
\frac{1-t}{3} - \phi _{_L} ^2 & = & {2 \over 5} \varepsilon - \frac{\varepsilon
^2 }{50} - \frac{3}{500} \varepsilon ^3 \nonumber
 \end{eqnarray} 

\noindent
Note that at the tricritical point, the slopes of the two phase
transition lines given by Eq.(\ref{slope}) and Eq.(\ref{pente}) are
equal as has been noticed by the original mean-field work of Ref. \cite{BEG}.

%\newpage
%%%%%%%%%%%%%%%%%%%%%%%%%%%%%%%%%%%%%%%%%%%%%%%%%%%%%%%%%%%%%%%
\section{The Lamellar-Disorder interfacial profile}
%%%%%%%%%%%%%%%%%%%%%%%%%%%%%%%%%%%%%%%%%%%%%%%%%%%%%%%%%%%%%%%
\noindent

In a simple
condensation transition, the interface width between the two homogeneous
phases, $\xi\sim t^{-1/2}$, has the same scaling behavior as the
typical
concentration fluctuations of the homogeneous phase.  In a modulated
phase, a second characteristic length scale exists and has to be
considered together with this correlation length. This is the wave-length
of the super structure, $q^*$.  In order to compute analytically the
profile of the order parameter for the disorder-lamellar interface close
to the tricritical point, where the two length scales are expected to be
uncorrelated, we propose the following ansatz~:

 \begin{equation} 
\phi(x)=\phi_{+} + \delta \phi(x) \;\;~~~~~~
\hbox {where}\;\;~~~~~~
\delta \phi(x)  = 
{\phi_{-} } f(x) + \phi_q  g(x) \cos(q^* x)
\label {ansatz}
 \end{equation}

Far from the interface, in the disordered region, 
$f(x) = 1$ and $g(x) = 0$, whereas in the lamellar phase region,
 $f(x) = -1$ and $g(x) = 2$. These are the boundary conditions for
 the differential equations of the interfacial profile.
The slowly spatially varying functions $f(x)$ and $g(x)$ play the
role of the two order parameters. The difference in the scaling
of $\phi_{-}$ and $\phi _q$ with $\varepsilon$ close to the tricritical
point, enables us to decouple the behavior of these two functions.
%\footnote{asymmetry at tcp}

Expanding the chemical potential up to the third order in
${\phi_{-} }$, we find~: 

\begin{equation}
\mu -t{ \phi_{+ }} - \phi_{+}^3 = \alpha {\phi_{-} } +3
{\phi_{+ }}\phi_{-}^2 +\phi_{-}^3
\end{equation}

\noindent
where $\alpha$ is define as:  
\begin{equation}
 \alpha \equiv 
\frac{ \partial^2 F_{_D} }{ \partial \phi_{_D}^2} = t 
+ 3 \phi_{+}^2 = 1- \frac{3}{5} \varepsilon + 
\frac {27} {100} \varepsilon^2 - {\varepsilon^3 \over 6}  
\end{equation}

Incorporating ansatz (\ref{ansatz}) into
$G_{_L}=F_{_L} - \mu \phi_{_L}$, we get~:  
\begin{eqnarray}
 F_{_L} -\mu \phi_{_L} = G(\phi_{+})+{ \alpha \over 2} (\delta \phi)^2 + {
\phi_{+}} 
(\delta\phi)^3  
 +{1 \over 4} (\delta\phi)^4 -(\alpha {\phi_{-} }+3 { \phi_{+} \phi_{-}^2}
+ \phi_{-}^3 ) \delta\phi +
 \\
 2 \{f''{\phi_{-} }+ 
\phi _q [(g''-(q^*)^2 g)\cos( q^* x)-2q^*g' \sin(q^*x) ]^2 -2(f'{\phi_{-} } +
 \phi _q (g' \cos( q^* x)-qg \sin(q^*x)) \}^2 \nonumber  
\end{eqnarray}

\noindent
 Close to the tricritical point, the two different length scales behave
differently. The profile is characterized by the width of the interface,
$\xi$ , which should diverge at $t=t_{tcp}$. However, in the same limit,
the modulation wavelength, $2\pi / q^*$, remains unchanged, as can be
checked from the numerical calculations (see Fig. 5).  Therefore, taking
the average of $F_{_L}-\mu \phi_{_{L}}$ over one period, yields~: 
 \begin{eqnarray} 
\langle F_{_L} -\mu \phi_{_L}\rangle&=&\nonumber
\\
G(\phi_{+})+2\phi_{-}^2 [(f'')^2 -(f')^2 ] &+&
 \phi_q^2 [ (g''-{1 \over 2} g)^2 +{g'}^2-\frac{1}{2}g^2] + 
{\alpha \over 2}\phi_{-}^2 f^2 +{\alpha \over 4}\phi_q^2 g^2 +
\\
{ \phi_{+}} \phi_{-}^3 f^3 + {3 \over 2}\phi_q^2 
{\phi_{-} }{ \phi_{+}}f g^2 &+& 
{1 \over 4 } \phi_{-}^4 f^4 +
{3\over 4}\phi _q^2 g^2 f^2 \phi_{-}^2 +{3\over 32}\phi _q^4 g^4-
(\alpha {\phi_{-} }+3{\phi_{+}}\phi_{-}^2 ){\phi_{-} }f \nonumber
\end{eqnarray}

\noindent
The variational principle with respect to $f(x)$ leads to the
relation~:

 \begin{equation}
 \alpha f +3{ \phi_{+}}
{\phi_{-} } f^2 + {3 \over 2}{\phi _q ^2 \over {\phi_{-} }}
{ \phi_{+}} g^2 + {3 \over 2} \phi _q ^2 f g^2 
+\phi_{-}^2 f^3 =  \alpha +3{ \phi_{+}}\phi_{-}
 \end{equation} 

Using expressions (\ref{expansion}) for $\phi_{+}$, 
$ \phi_{-}$ and $ \phi _q $, we get up to the first order in $\varepsilon$: 

 \begin{equation}
 f(x)=1- {1 \over 2} g^2 + {9 \over 10}  g^2 ({g^2 \over 4}-1)\varepsilon
\label{fonction}
\end{equation}

\noindent
This relation enables us to find, when applying the variational
principle with respect to $g(x)$, the following differential equation~:
\begin{equation}
4g{''}= {{\alpha -1 } \over 2 } g + 3 f g {\phi_{-} }{ \phi_{+}}
 + {3 \over 2}\phi_{-}^2 f^2 g + {3 \over 8 } \phi _q ^2 g ^3 
\end{equation}

\noindent
which can be rewritten at the lowest order in $\varepsilon$ as~: 

\begin{equation} 
{20 \over 3}g{''}(x)= (g - g^3 +{3 \over16} g^5) \varepsilon^2
\end{equation} 

Taking into account the boundary conditions for $g(x)$,
the analytical solution of this differential equation is found to be~:

 \begin{equation} 
g(U) = \frac{ 2{\rm e}^{-U}  }{\sqrt{1 + {\rm e}^{-2U}}}=
       \frac{2}{ \sqrt{1 + {\rm e}^{2U}}}
 \end{equation} 

\noindent
where $U(x)=\frac{1}{2}\sqrt{3 \over 5} ( {x-x_0})\,\varepsilon$, and
$x_0$ can be set to zero without lost of generality. 
Using Eq. (\ref{fonction}) we can now obtain for $f$
\begin{equation}
f(U)= \tanh (U) - {9 \over 10} ~{\cosh}^{-2} (U)~\varepsilon
\end{equation}
 
The two functions $f(x)$ and $g(x)$ 
determine the interfacial profile $\phi(x)$ close to the tricritical
point
and are plotted in Fig. 3 and Fig. 4.  
The width of the interface scales as $\xi \sim \varepsilon^{-1} $.
Namely, the correlation length exponent $\nu = 1 $
is recovered, in agreement with the
scaling laws associated with the tricritical point \cite{REV}.
We note that the main reason it is possible to decouple the 
differential equations for $f$ and $g$ (to lowest order in $\varepsilon$)
is due to the different scaling of $\phi_q$ and $\phi_{-}$ with respect
to $\varepsilon$ close to the tricritical point [see Eq. (\ref{expansion})].

%\newpage
%%%%%%%%%%%%%%%%%%%%%%%%%%%%%%%%%%%%%%%%%%%%%%%%%%%%%%%%%%%%%%%
\section{The Disorder-Lamellar interfacial tension}
%%%%%%%%%%%%%%%%%%%%%%%%%%%%%%%%%%%%%%%%%%%%%%%%%%%%%%%%%%%%%%%
\noindent

The interfacial tension $\gamma$, is
defined as the excess of free energy due to the interface.
It can be written as~:

 \begin{equation} 
\gamma =\int \Gamma(x) dx =\int [G - \frac{G_{_D} + G_{_L} } {2}] dx = 
\int [F - \frac{F_{_D} + F_{_L} } {2} - \mu \,\delta \phi(x) ] dx
\label{surf}
\end{equation} 

Since up to second order in $\varepsilon$, the even
part of $ \Gamma(x)$ vanishes, we have to keep terms in $\Gamma$ up to
third order  in $\varepsilon$.  
\begin{eqnarray}
\Gamma =2\phi _q^2 (g')^2 + \frac{\alpha -1}{4} \phi_q^2 (g^2 - 2)
+ 3 \phi_q^2 {\phi_{-} } \phi_{+} ( \frac {f g^2}{2} + 1 )  +
\frac {3}{2} \phi _q ^2 \phi_{-}^2 ( \frac {f^2 g^2}{2} -1) \nonumber
\\ 
+ \frac{3}{4} \phi_q^4 ({g^4 \over 8} -1)  - \phi_{-}^2
( \alpha + 3{ \phi_{+}}{\phi_{-} })f + \frac{\alpha}{2} \phi_{-}^2
(f^2 -1)+\phi_{-}^3 \phi_{+}f^3
\end{eqnarray}

\noindent
This expression can be written as function of $f(x)$ only~:
\begin{equation}
 \Gamma =(1- f^2)({3 \over{50}} - \frac{\alpha}{2}\phi_{-}^2  +
3 \phi _q^2 {\phi_{-} }\phi_{+}-\frac {3}{2}\phi _q^2 
\phi_{-}^2 - \frac{3}{8} \phi_q^4 )+ \hbox{odd terms} 
\end{equation}

\noindent
and to third order in $\varepsilon$
\begin{equation}
\Gamma = \Bigl( {27 \over 250} {\cosh ^{-2}}(U) + 
\hbox{odd terms} \Bigr)\varepsilon^3  + O(\varepsilon ^4) 
\end{equation}

\noindent
where the odd terms in the above equations are odd 
with respect to $x \to -x$ and will vanish upon integration over $x$, yielding
\begin{equation}
\gamma = {18 \over 25 } \sqrt{3 \over 5} \varepsilon ^2
\end{equation}

\noindent
This result corresponds to the
expected  tricritical exponent $\mu = 2$, in agreement
with the hyperscaling relation $ \mu= (d-1) \nu$, as the upper
critical dimension for the tricritical point is $d=3$ \cite{REV}.

We note that in order to have the full expression of $f(x)$ up to first
order in $\varepsilon $,  one needs to expand $g(x)$ up to first
order in $\varepsilon$.  Indeed, if $g(x)= g_o (x)+g_1(x)\varepsilon $, then
$f(x)$ can be written as~: 

\begin{equation}
 f(x)=1- {1 \over 2} g^2 +{9 \over 10} g^2 ({g^2 \over 4}-1)\varepsilon  -
g_o (x) g_1(x) = f_o (x)+ f_1(x)\varepsilon. 
\end{equation} 

\noindent
Nevertheless, only the leading order terms of $g(x)$ and
$f(x)$ are required to compute the interfacial tension. 

Numerical studies of this  one-dimensional 
interface without assuming single-mode
dominance give results in good agreement
with the analytical calculation presented above. 
We have discretized the order
parameter $\phi$ and chosen 
the grid spacing to be much smaller than
 $1/ q^*$
(typically 180 grid points per period). The free
energy (\ref{free}) can then be minimized 
using the conjugate gradient method.

The free energies for the disordered (isotropic)
and lamellar phases (without using the single mode approximation)
are computed
in order to determine the chemical potential for the
coexisting phases. Then, we compute the free energy associated with the
interface, in an interval much bigger than the expected interface length,
in order to neglected boundary effect. The boundary conditions
are taken as reflecting (Dirichlet)
boundary conditions.
The same procedure of Ref. \cite{netz} using Eq. (\ref{surf})
is employed to calculate the interfacial 
energies. For small $\varepsilon=t_{tcp}-t$, the optimum
wave vector is still $q^*= 1/ \sqrt{2}$, within a good precision.
The profiles of the interface for
$\varepsilon=t_{tcp}-t=$0.35 and $\varepsilon=$0.03 
are shown in Figs. 5 and 6, respectively. 

The interfacial tension 
is plotted on Fig. 6 as a function of $ t_{tcp} -
t$. The log-log plot in 6a as well as the ${\gamma^{1/2}}$ plot
in 6b manifests clearly that our analytical result are: (i) in good
agreement with the numerical results obtained from the full minimization
of the interfacial free energy without assuming the single mode dominance;
and (ii) that indeed we obtained the expected scaling behavior of the
tricritical point. 
 
%%%%%%%%%%%%%%%%%%%%%%%%%%%%%%%%%%%%%%%%%%%
\section{Discussion and Conclusions}
%%%%%%%%%%%%%%%%%%%%%%%%%%%%%%%%%%%%%%%

 In this paper we have derived analytical expressions for the interface
profiles between a simple modulated phase with a lamellar symmetry and a
disordered one. Our results are valid close to the tricritical point since
we assume that the modulated phase is characterized by its dominant
$q$-mode in the bulk.
 This assumption is found to be compatible with the scaling laws close to
the tricritical point. Our analytical (but approximated) results are in
accord with numerical minimization of the full free energy functional
having the same one-dimensional symmetry.

Unlike simple interfaces between two coexisting homogeneous phases, here
the interface depends on two separate length scales. Close to the
tricritical point, the interface width $\xi$ diverges and hence gets
decoupled from the periodicity of the lamellar phase. It is interesting to
note that in this case, the interfacial region can involve quite a number
of lamellae periods as is shown in Figs. 4 and 5. For $\varepsilon=$0.03 the
interfacial region spans over 100 lamellae. 

Our results should be compared with
the ones obtained
by Fredrickson and Binder\cite{FB} since they obtained 
a similar envelope equation for the profile (eq.~(27) for $g$
in our notation). In Ref. \cite{FB} the metastable interface
close to a symmetric critical point is considered.
Due to the symmetry,
they have $f=0$ (in our notation). The high order terms ($\phi^5$)
in their profile equation arises from the first-loop correction
(order $\phi^6$) in their mean-field effective Hamiltonian. In our case,
due to the asymmetry, the coupling of
$g$ to $f\ne 0$ generates the high order terms 
even within mean field theory. Needless to say that the scaling
of the correlation length and surface tension is different in the
two models because of the difference between critical and tricritical
behavior.

It will be interesting to try to obtain experimental data on lamellar to
disorder interfaces close to the tricritical point. Unfortunately, in many
experimental realizations the one-dimensional phase diagram studied here
is modified. The lamellar to disorder transition becomes metastable
since it is pre-empted close to the tricritical point
by phases with hexagonal and cubic symmetry \cite{netz}.
Then, the lamellar-to-disordered interfaces can exist 
in thermodynamical equilibrium but only
far from the tricritical point. Close to the tricritical point
the lamellar to disordered transition describes only
a metastable situation which is of importance, for example,
for systems with slow dynamics like copolymer mixtures \cite{FB,HS}.
We hope the systematic expansion we present can be applied to other
more complicated situations where it is also possible to decouple
the important length scales in the system.

\bigskip\bigskip
\noindent
{\it Acknowledgment}

\noindent
We would like to thank J. Beckh\"ofer, E. Meron, R. Netz, M. 
Schick and B. Widom for discussions and comments.  
In particular, we thank T. Garel for very useful remarks and for
bringing Ref. \cite{FB} to our attention.
This work was
partially supported by a grant from the United States-Israel Binational
Science Foundation under grant No. 94-00291. One of us (SVG) thanks the
French Ministry of Foreign Affairs for a research fellowship.

%\begin{figure}
%\epsfysize=20\baselineskip
%\centerline{\hbox{ \epsffile{diap.ps} }}
%\caption [Figure 1a : phase diagram]
%         {\small\protect 

%\end{figure}

%\begin{figure}
%\epsfysize=20\baselineskip
%\centerline{\hbox{ \epsffile{potchi.ps} }}
%\caption [{Figure 1b : phase diagram}]
%         {\small\protect Schematic 
%
%\end{figure}

%\begin{figure}
%\epsfysize=20\baselineskip
%\centerline{\hbox{ \epsffile{fg.ps} }}
%\caption [{Figure 1b : phase diagram}]
%         {\small\protect 
%
%\end{figure}
%
%\begin{figure}
%\epsfysize=20\baselineskip
%\centerline{\hbox{ \epsffile{profile.ps} }}
%\caption [{Figure 1b : phase diagram}]
%         {\small\protect Analytical interfacial profile for t=0.35.}
%\end{figure}
%
% 
%\begin{figure}
%\epsfysize=20\baselineskip
%\centerline{\hbox{ \epsffile{dessin1.ps} }}
%%\caption [Figure 1b : dessin1.ps]
%         {\small\protect 
%
%\end{figure}  

%\begin{figure}
%\epsfysize=20\baselineskip
%\centerline{\hbox{ \epsffile{dessin2.ps} }}
%\caption [{Figure 1b : dessin2.ps}]
%         {\small\protect 
%\end{figure}  

%\begin{figure}
%\epsfysize=20\baselineskip
%\centerline{\hbox{ \epsffile{ech1.ps} }}
%\caption [{Figure 1b : phase diagram}]
%         {\small\protect 
%\end{figure}  

%\begin{figure}
%\epsfysize=20\baselineskip
%\centerline{\hbox{ \epsffile{ech2.ps} }}
%\caption [{Figure 1b : phase diagram}]
%         {\small\protect 
%\end{figure}  

\newpage
\section*{Figure Captions}

\noindent{\bf Fig. 1}~~~
The calculated phase diagram from the modulated phase free energy $F_{_L}$
as function of the reduced temperature - average order
parameter ($t$, $\phi_o$). The phase diagram is symmetric with
respect to $\phi_o \to -\phi_o$. 
The two symmetric
tricritical points are denoted by {\it tcp}. Below the $tcp$ 
temperature,
there are two regions of coexistence of the lamellar phase and
disordered phase (L+D$_1$) and (L+D$_2$). 
At even lower temperatures, 
the lamellar phase disappears. We
then recover the usual coexistence region
between two different disordered phases: D$_1$+D$_2$~ \cite{DA}.

\bigskip
\noindent{\bf Fig. 2}~~~
Same as in Fig. 1 but as function of
the chemical potential - reduced temperature ($\mu$, $t$). 
The continuous line is a second-order transition line. 
The dashed line is a first-order one. 
They meet at two tricritical points ({\it tcp}). The two transition lines
separate 
the lamellar phase region from the disordered region. The phase diagram
is symmetric with respect to $\mu \to -\mu$. The $\mu=0$ 
first order transition line meets the two other first order
lines at a triple point where all three phases: L, D$_1$ and D$_2$
coexist.

\bigskip
\noindent{\bf Fig. 3}~~~
(a) Analytical profiles 
of the two functions $f(x)$ and $g(x)$ for 
$\varepsilon$ = $t_{tcp}-t$ = 0.35.
The plot scans the entire interfacial region between
a bulk lamellar phase (left side where $g\to 2$, $f \to -1$) 
and the disordered phase (right side where $g \to 0$, $f \to 1$).
(b)
Analytical interfacial profile $\phi(x)$ using the same functions
$f(x)$ and $g(x)$ of (a).
The value of the average concentration in the two coexisting
phases is: $\phi_0=0.25$ for the lamellar phase and $\phi_0=0.50$ for
the disordered one.

\bigskip
\noindent{\bf Fig. 4}~~~
Numerical interfacial profile for $\varepsilon=0.35$ and
$\mu=0.322$. In order to demonstrate that 
the most dominant bulk mode remains
unchanged even in the interfacial region, 
the 
bulk lamellar phase 
is plotted as 
a dotted line. The horizontal line just below
$\phi_0=0.5$ denotes the
average concentration of the bulk disordered phase.
\bigskip

\bigskip
\noindent{\bf Fig. 5}~~~
Numerical interfacial profile for $\varepsilon=0.03$ and
$\mu=0.249$. 
The horizontal line just above
$\phi_0=0.3$ denotes the
average concentration of the bulk disordered phase.
 The width of the interface is much larger than  $2\pi / q^*$,
and over 100 lamellae are located in this interfacial region.

\bigskip
\noindent{\bf Fig. 6}~~~
(a) Interfacial tension as function of $\varepsilon=t_{tcp}-t$
plotted on a log-log plot.
The straight line is the analytical interfacial
tension $\gamma$. The dots represent the numerical
results as is discussed in the text. The linear
dependence of both the  analytical and numerical results
are in accord with the general scaling behavior expected close to the
tricritical point.
 (b) Same results as in (a) but plotted differently. The square root of
the interfacial tension $\gamma$ is plotted as a function of
$\varepsilon=t_{tcp}-t$. The linear dependence is another confirmation of the
tricritical point scaling $\gamma \sim \varepsilon^2$


\begin{thebibliography}{99}


 
\bibitem{SC} M. Seul and D. Andelman, Science {\bf 267}, 476 (1995).

\bibitem{netz} R. R. Netz, D. Andelman and M. Schick, 
{Phys. Rev. Lett.} {\bf 79}, 1058 (1997).

\bibitem{FB} G. H. Fredrickson and K. Binder, 
J. Chem. Phys. {\bf 91}, 7265
(1989).


\bibitem{BR} A. Brazovskii, Zh. Eksp. Teor. Fiz. {\bf 68}, 175 (1975);
[Sov. Phys. JETP {\bf 41}, 85, (1975)].

\bibitem{GA} T. Garel and S. Doniach, Phys. Rev. B  {\bf 26}, 325 (1982).

\bibitem{GP} G. Grinstein and R. A. Pelcovits, Phys. Rev. A {\bf 26},
 915 (1982). 
% not very relevant

\bibitem{BEG} M. Blume, V.J. Emery and R.B. Griffiths, Phys. Rev. A {\bf
4}, 1071, (1971).  

\bibitem{DA} D. Andelman, F. Brochard and J.-F. Joanny, J. Chem. Phys.
{\bf 86}, 3673 (1987). 

\bibitem{REV} I.D. Lawrie and S. Sarbach,``{\sl Theory of Tricritical Points"}
in {\it Phase Transitions and Critical Phenomena}, Vol. 9, C. Domb and
M.S. Green editors, (McGraw-Hill, New York). 

\bibitem{HS} P. C. Hohenberg and J. B. Swift, Phys. Rev. E {\bf 52},
 1828 (1995).



%\bibitem{malomed} B. A. Malomed, A. A. Nepomnyashchy, M. I. Tribelsky,
%Phys. Rev. A {\bf 42}, 7244 (1990). 
% domains boundaries in convection patterns
%hexagonal domains close to walls.

%\bibitem{newell} A. C. Newell, T. Passot, C. Bowman, N. Ercolani and
% R. Indik, Physica D {\bf 97}, 185 (1996).
%Defects in diffusion type equations.
 
\end{thebibliography}
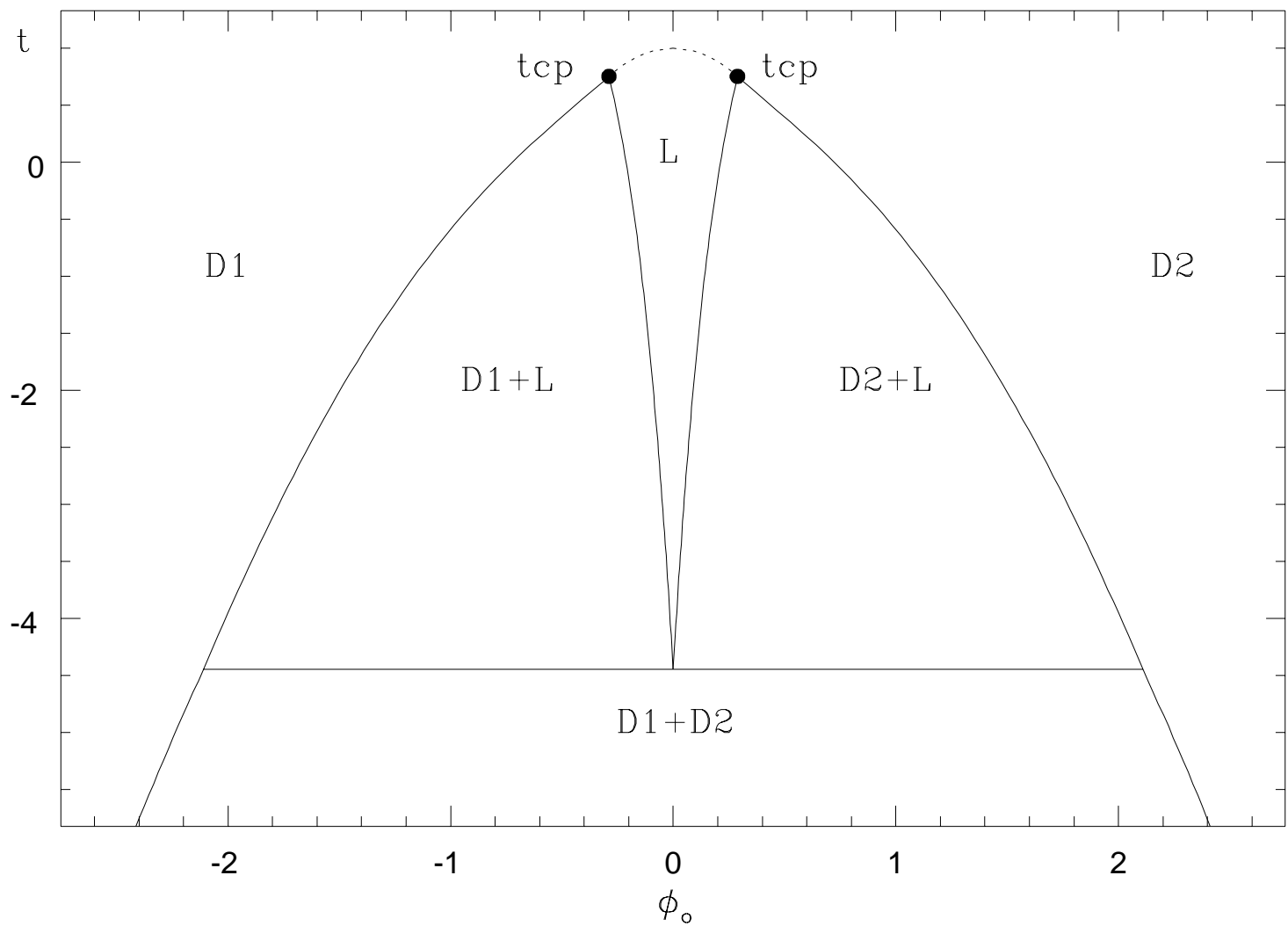
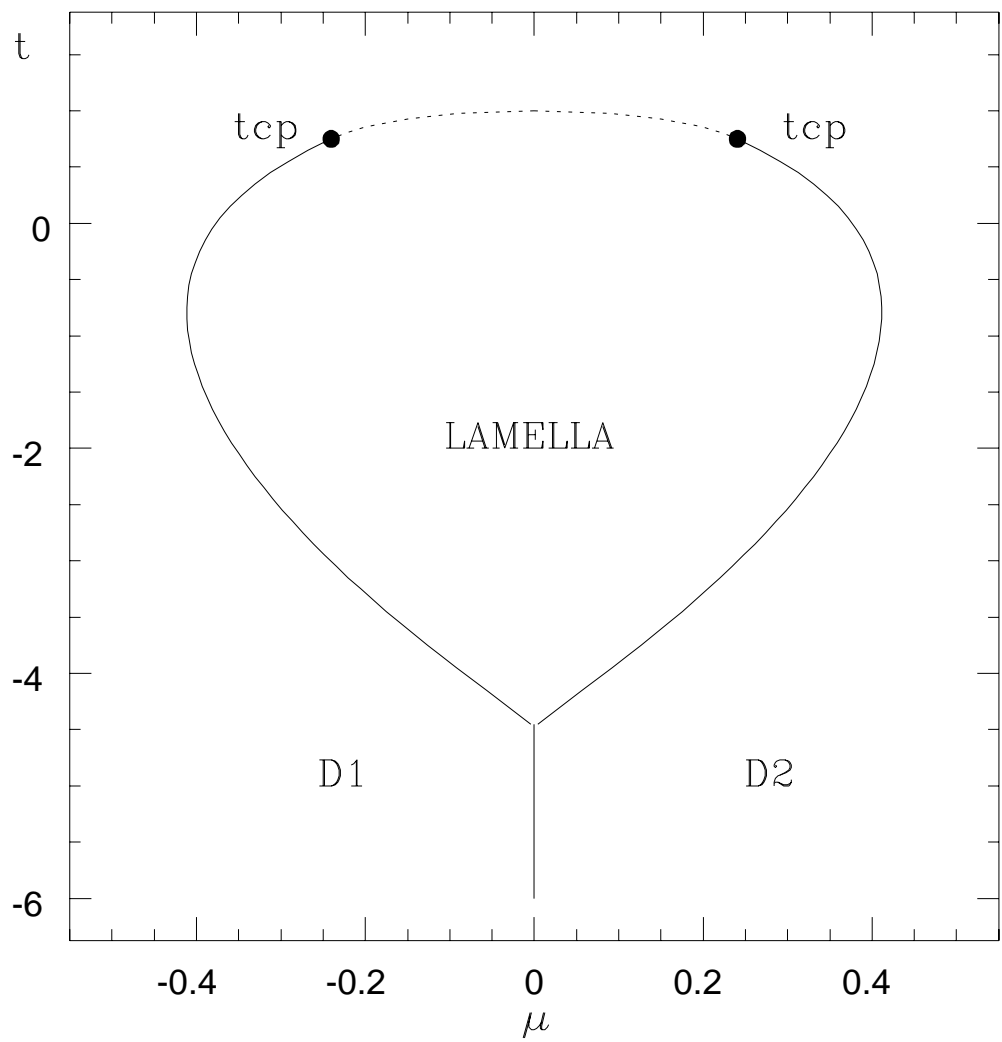
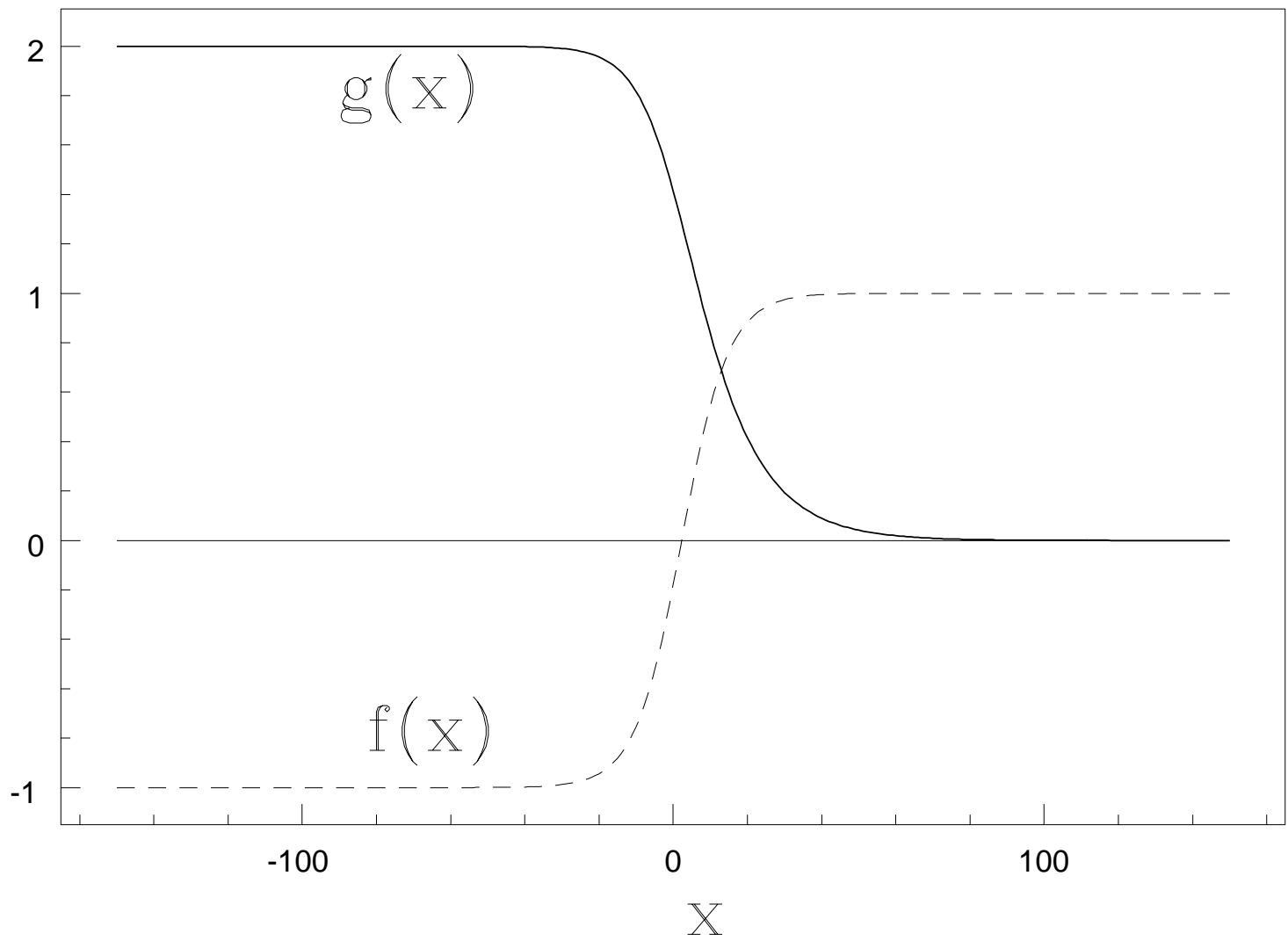
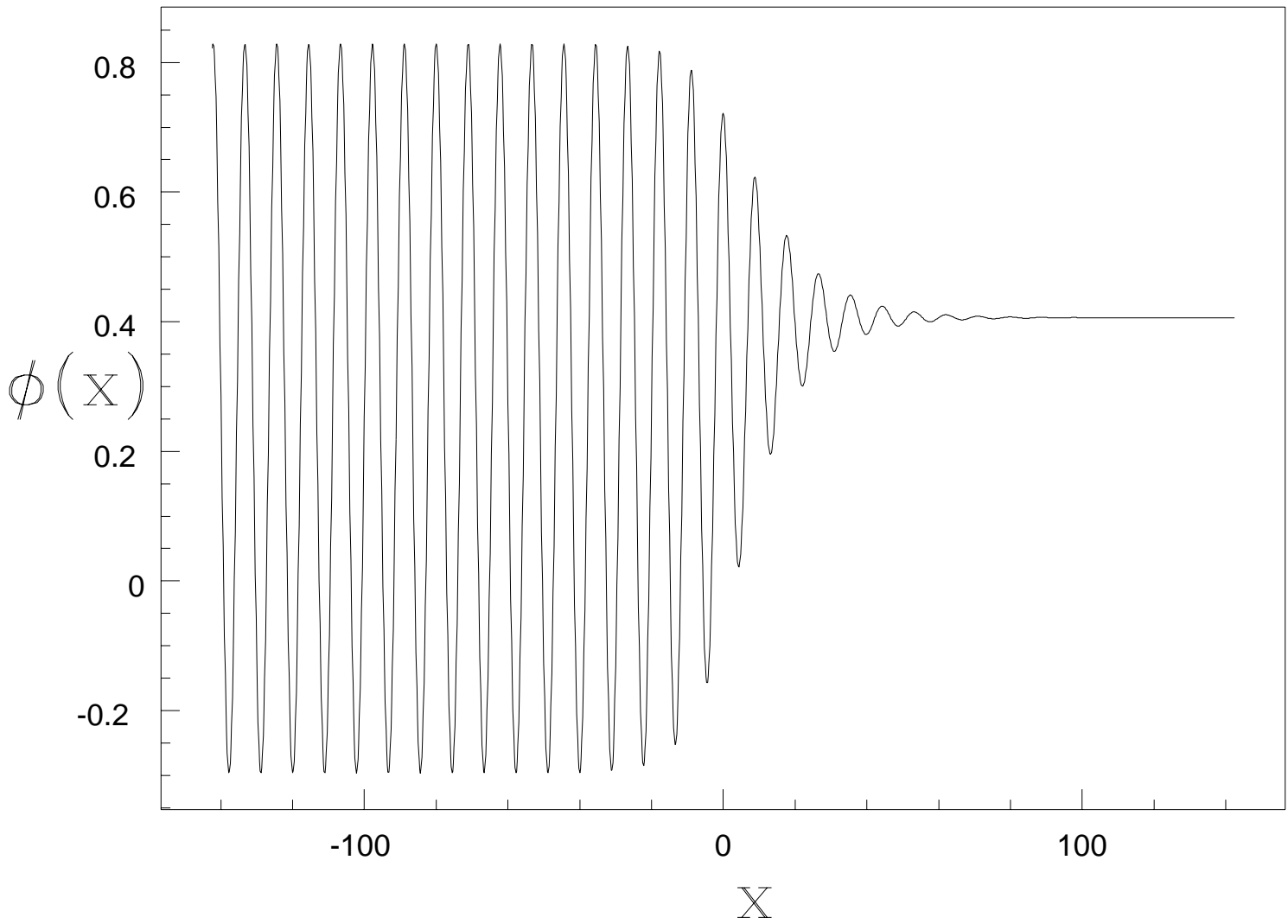
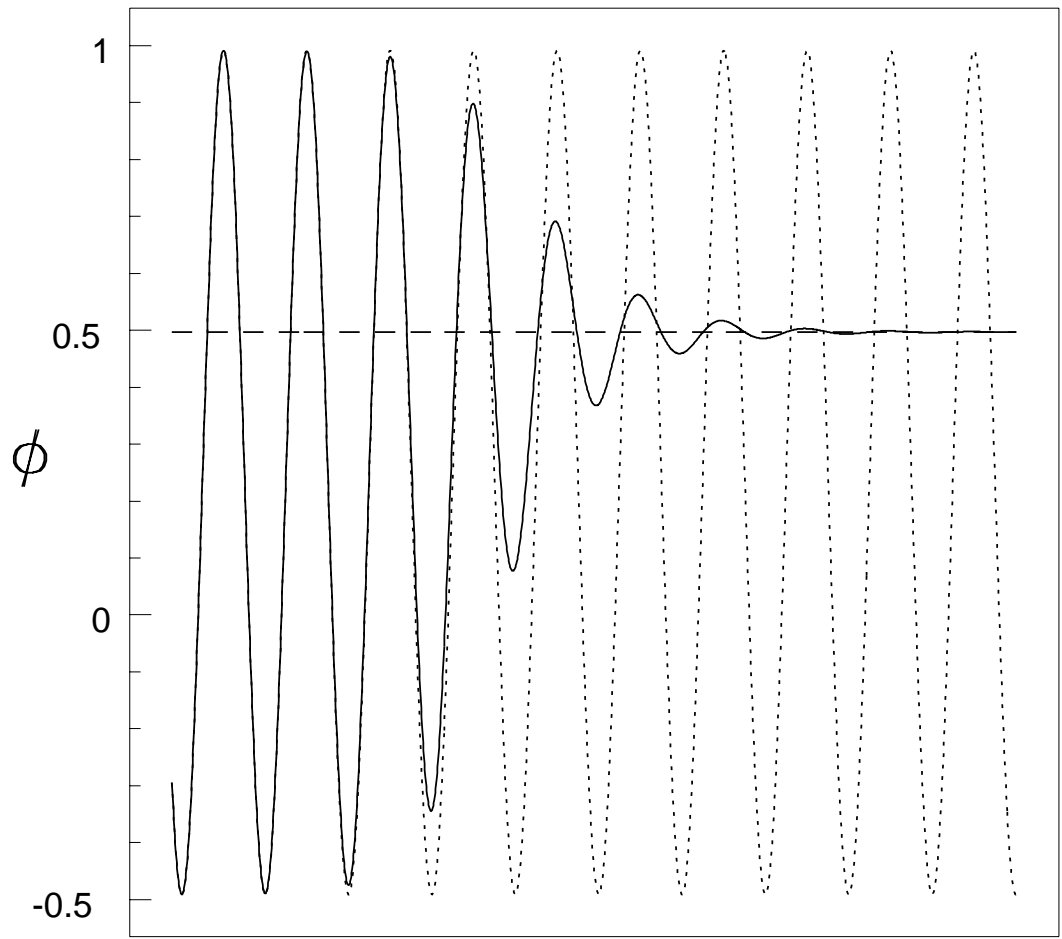
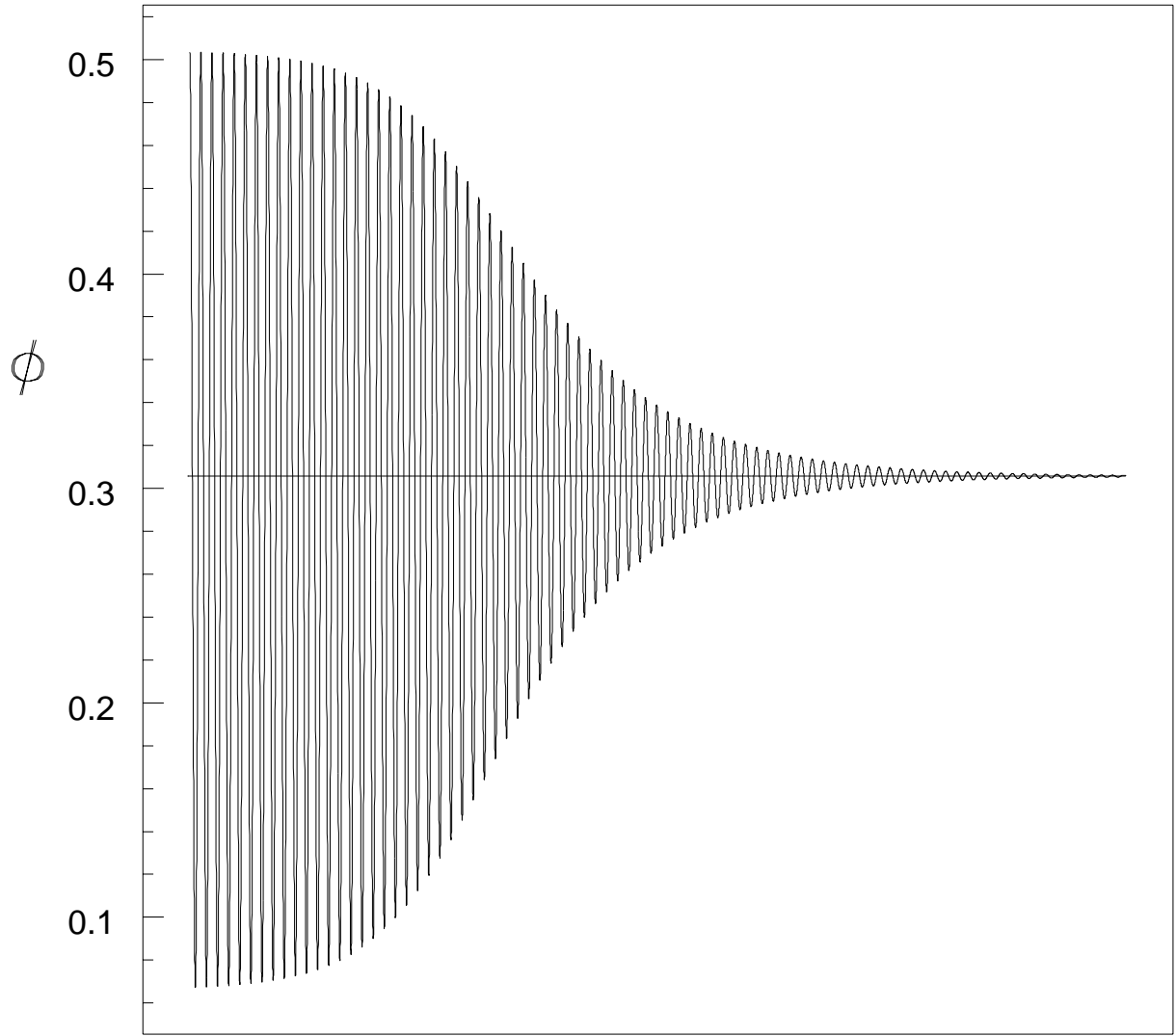
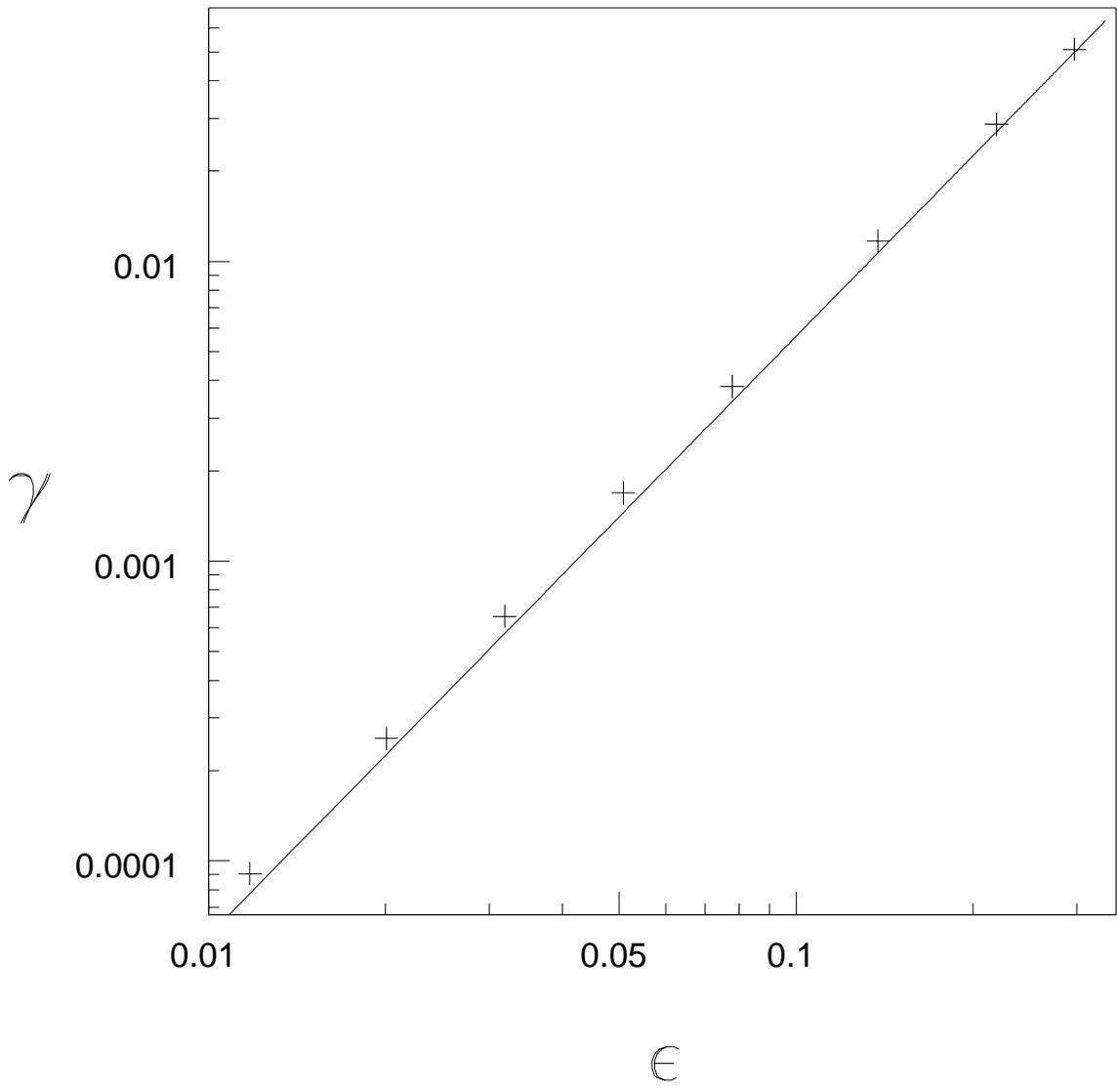
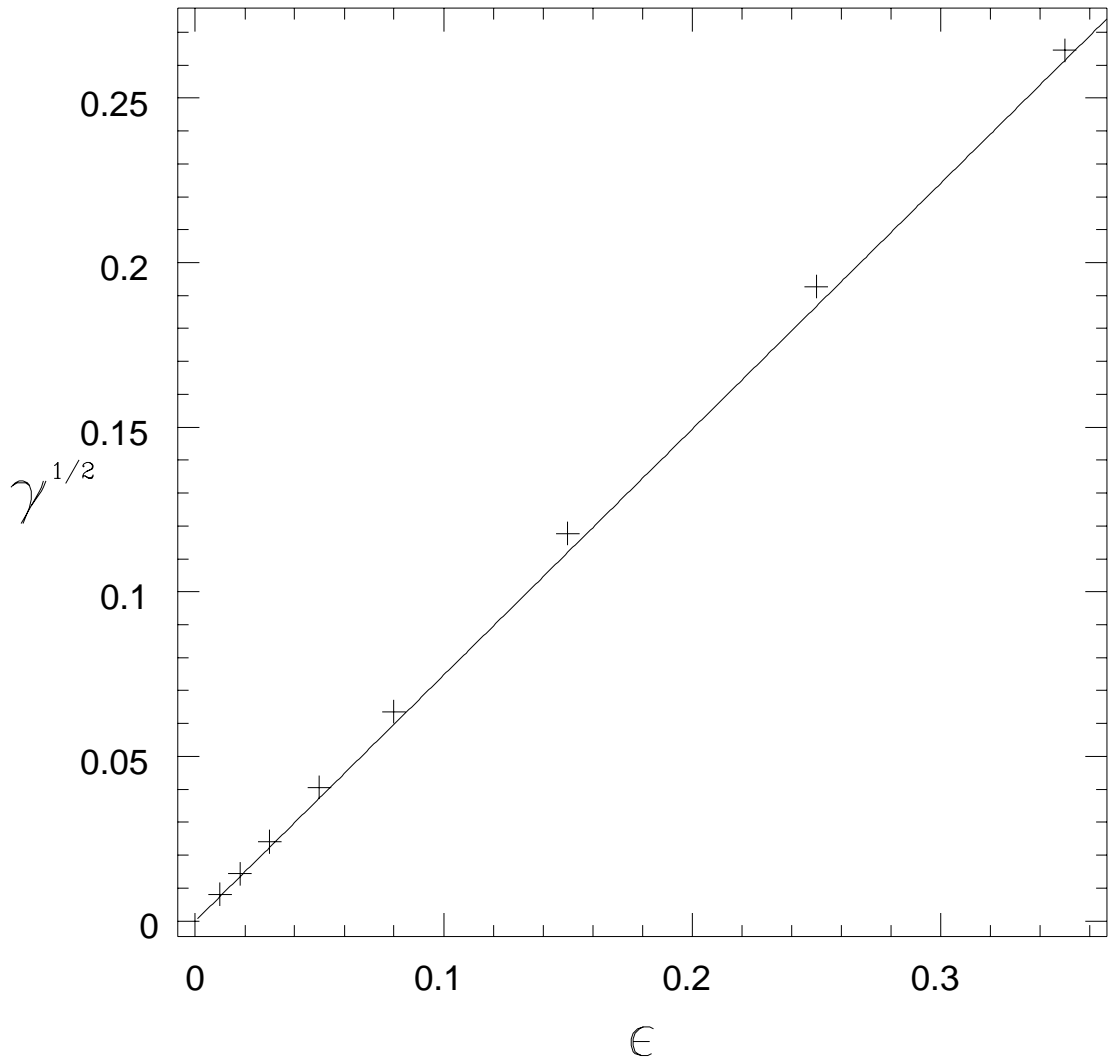
\end{document}